\documentclass[reprint,showpacs,amssymb,amsmath,aps,pra]{revtex4-1}

\usepackage{graphics}
\usepackage{bm}
\begin{document}
\title{Violation of majorization relations in entangled states
and its detection by means of generalized entropic forms}
\author{R. Rossignoli, N. Canosa}
\affiliation{Departamento de F\'{\i}sica, Universidad Nacional de La Plata,
C.C.67, La Plata (1900), Argentina}
\begin{abstract}
We examine the violation of the majorization relations between the eigenvalues
of the full and reduced density operators of entangled states of composite
systems and its detection using generalized entropic forms based on arbitrary
concave functions. It is shown that the violation of these relations may not
always be detected by the conditional von Neumann and Tsallis entropies (for
any $q>0$). Families of smooth entropic forms which are always able to detect
such violations are, however, provided. These features are then examined for
particular sets of mixed states in a two-qudit system, which for $d\geq 3$ may
exhibit different types of violation of the majorization relations. Comparison
with the Peres criterion for separability is also shown.
\end{abstract}
 \maketitle

Entanglement is now recognized as one of the most important features of quantum
mechanics. It represents the possibility of composite quantum systems to
exhibit correlations which cannot be accounted for classically. It plays
therefore a fundamental role both in philosophical problems concerning the
interpretation of quantum theory \cite{S.35} as well as in applications like
quantum computation, quantum cryptography and quantum teleportation 
\cite{Di.95,NC.01,GM.02,Ek.91,B.93,B.96}, where it is actually viewed as a 
resource. Formally, a quantum system composed of two subsystems $A$ and $B$ is 
said to be in a separable or unentangled state if the density operator of the 
whole system can be written as a convex combination of uncorrelated densities 
\cite{W.89},
 \begin{equation}
 \rho=\sum_\alpha\omega_\alpha\rho^\alpha_A\otimes\rho^\alpha_B,\;\;
  0\leq\omega_\alpha\leq 1\,,
 \end{equation}
where $\sum_{\alpha}\omega_\alpha=1$ and $\rho^\alpha_{A}$, $\rho^\alpha_B$ are
density operators for each subsystem.  Otherwise, the system is said to be {\it
entangled} or inseparable. When separable, the system fulfills all Bell
inequalities and satisfies other properties characteristic of classical
composite systems, making it unsuitable for the previous applications. 
Pure states ($\rho=|\Psi\rangle\langle\Psi|$) are separable only for tensor
product states ($|\Psi\rangle=|\phi_A\rangle|\phi_B\rangle$), but in the case
of mixed states it is in general much more difficult to determine whether a
given density is separable or not. There are several simple necessary
conditions for separability \cite{P.96,HHH.96,HH.96,HH.99,CAG.99,NK.01} but 
they are not sufficient in general.

Among them, one is directly related with disorder and entropy \cite{NK.01}. Let
\[ \rho_A={\rm Tr}_B[\rho],\;\;\rho_B={\rm Tr}_A[\rho]\,,\]
be the reduced density matrices of the subsystems, where ${\rm Tr}_{A,B}$
denotes partial trace. It has been shown in \cite{NK.01} that if $\rho$ is
separable, then $\rho$ is {\it more mixed} (or disordered) than both $\rho_A$
and $\rho_B$ (the disorder criterion for separability). This property, which
can be written as $\rho\prec\rho_A$, $\rho\prec\rho_B$ means in the present
situation that
 \begin{equation}
{\cal S}_i\equiv\sum_{j=1}^ip_i\leq {\cal S}^A_i\equiv\sum_{j=1}^ip^A_i,
\;\;{\rm for}\;\;i=1,\ldots,d_A\,,\label{m}
\end{equation}
(and similarly ${\cal S}_i\leq {\cal S}_i^B$ for $i=1,\ldots,d_B$)  where
$p_i$, $p_i^{A}$ are, respectively, the eigenvalues of $\rho$ and $\rho_A$
sorted in {\it decreasing} order and $d_A$ is the dimension of $\rho_A$. The
set of eigenvalues of $\rho$ is then {\it majorized} by those of $\rho_A$ and
$\rho_B$. For a pure state, Eqs.\ (\ref{m}) can be fulfilled for $A$ and $B$
only if $p_1^A=p_1^B=1$, implying that the state must be a tensor product
state. However, for mixed states the disorder criterion is only a necessary
condition for separability. No sufficient condition for mixed states can
actually be based on the knowledge of the eigenvalues of $\rho$ and
$\rho_{A,B}$ alone \cite{NK.01,VW.01}. Nevertheless, it is the strongest
spectral criterion and expresses a fundamental classical property of separable
systems: a classical bipartite system is always more disordered than its
subsystems, as the set of joint probabilities $\{p_{ij}\}$ are always majorized
by the set of marginal probabilities $\{p^A_{i}=\sum_jp_{ij}\}$.

The disorder criterion admits a simple entropic formulation \cite{RC.02}.
Let us define the general entropic forms \cite{CR.02}
 \begin{equation}
S_f(\rho)={\rm Tr}f(\rho)\,,\label{Sf}
 \end{equation}
where $f$ is an arbitrary smooth concave function satisfying $f(0)=f(1)=0$.
These forms satisfy most basic properties of the conventional entropy except
those related with additivity \cite{CR.02}. In particular, if $\rho\prec\rho'$,
then $S_f(\rho)\geq S_f(\rho')$ for any $f$ of the previous form
\cite{RC.02,W.78}. The disorder criterion implies then that in a separable
state, the generalized entropy of the whole system is not less than those of
the subsystems:
 \begin{equation}
S_f(\rho)\geq S_f(\rho_A)\,,\label{Sf1}
  \end{equation}
and similarly, $S_f(\rho)\geq S_f(\rho_B)$. In particular, the standard von
Neumann entropy
 \begin{equation}
S(\rho)=-{\rm Tr}\rho\ln\rho\,,\label{S}
 \end{equation}
corresponds to $f(\rho)=-\rho\ln\rho$, and  Eq.\ (\ref{Sf1}) implies that the
von Neumann conditional entropy $S(\rho)-S(\rho_A)$ \cite{W.78} is non-negative
in any separable state \cite{HH.96}. Similarly, the Tsallis entropy \cite{T.88}
\begin{equation}
S_q(\rho)=(1-{\rm Tr}\rho^q)/(q-1)\,,\label{Sq}
\end{equation}
corresponds to $f(\rho)=(\rho-\rho^q)/(q-1)$, which is concave for all $q>0$,
and Eq.\ (\ref{m}) implies that the Tsallis conditional entropy \cite{AR.01}
$[S_q(\rho)-S_q(\rho_A)]/{\rm Tr}\rho_A^q$ is non-negative in any separable
state. For $q\rightarrow 1$, $S_q(\rho)\rightarrow S(\rho)$.

For a given choice of $f$, Eq.\ (\ref{Sf1}) provides therefore a necessary test
for separability of mixed states. In the von Neumann case, the criterion is
actually rather weak, but in the Tsallis case the validity of Eq.\ (\ref{Sf1})
for all $q>0$ provides a more stringent requirement \cite{AR.01,TLB.01,TPA.02},
which has been shown to yield sufficient conditions for separability for some
classes of states, like Werner states for $n$-qubits \cite{AR.01} as well as
any two-qubit state diagonal in the Bell basis \cite{RC.02}.

Let us come now to the central point of this work. It is also true that if
$S_f(\rho)\geq S_f(\rho')$ for {\it any} $f$ of the previous form, then
$\rho\prec\rho'$ \cite{W.78}, as will be explicitly shown below. Hence the
validity of Eq.\ (\ref{Sf1}) {\it for all} $f$ (and its partner for $\rho_B$)
provides an equivalent formulation of the disorder criterion. On the other
hand, its validity for a particular $f$ does not imply of course that
$\rho\prec\rho_A$. Moreover, {\it the validity of Eq.\ (\ref{Sf1}) for all
$q>0$ in the Tsallis case does not imply} $\rho\prec\rho_A$. This may happen
whenever the first violation of Eqs.\ (\ref{m}) occurs for $i>1$. For
sufficiently large $q$, $S_q(\rho)\approx (1-m_1p_1^q)/(q-1)$, where $m_1$ is
the multiplicity of the largest eigenvalue $p_1$ of $\rho$. Hence, if
$p_1>p_1^A$, $S_q(\rho)<S_q(\rho_A)$ for sufficiently large $q$. However, if
the first violation takes place for $i>1$, Eq.\ (\ref{Sf1}) may remain valid
for all $q>0$, as can be seen in simple examples. While in a two qubit system
the only possible breakdown of Eqs.\ (\ref{m}) is for $i=1$ (as ${\cal
S}_2^A=1$), this is not the case in systems of larger dimensionality, where it
is actually not rare to encounter entangled states in which the first violation
of Eqs.\ (\ref{m}) occurs  for any $i>1$, as shown in the example below. In
such cases the Tsallis conditional entropy may not detect entanglement even
when the disorder criterion does. The same will hold for any other set of
functions $f(p)$ whose maximum is obtained for $p\rightarrow 1$ in the limit
where it becomes sharp \cite{CR.02}.

We shall now provide a set of {\it smooth} concave functions which are {\it
always} able to detect the breakdown of the inequalities (\ref{m}), such that
$S_f(\rho)<S_f(\rho_A)$ for some $f$ of this set if Eqs.\ (\ref{m}) are
violated for some $i\geq 1$. This can be achieved with functions having a
maximum at an optimizable point $\alpha\in (0,1)$, which can be made 
arbitrarily sharp by adjusting a second parameter. A possibility is
 \begin{equation}
f(\rho)=g_t(\rho-\alpha)-(1-\rho)g_t(-\alpha)-\rho g_t(1-\alpha)\label{gt}\,,
 \end{equation}
where $\alpha\in[0,1]$ and $g_t(x)$ is a smooth concave function satisfying
 \[\lim_{t\rightarrow\infty}g_t(x)=-|x|/2\,.\]
An example is (see Fig.\ 1)

 \begin{figure}[t]
\vspace*{-3cm}

 \centerline{\scalebox{0.6}{\includegraphics{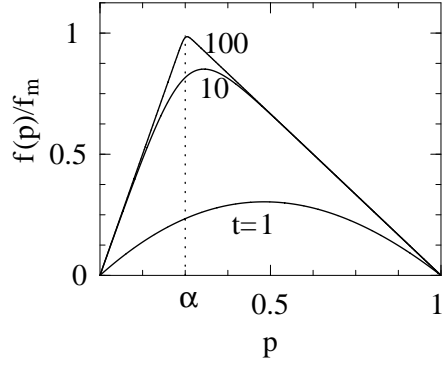}}}
 \vspace*{-9.5cm}

\caption{The entropic functions (\ref{gt}) with  $g$ given by Eq.\ (\ref{gtx}),
for $\alpha=1/4$ and different values of $t$. $f_m=\alpha(1-\alpha)$ denotes
the maximum value in the $t\rightarrow \infty$ limit.}
 \label{f1}\end{figure}

 \begin{equation}
 g_t(x)=-(2t)^{-1}\ln \cosh(tx)\,.\label{gtx}
 \end{equation}
For any of these functions, $f(0)=f(1)=0$, while for $t\rightarrow\infty$ and
$0<\alpha<1$,
\begin{equation}
f(p)\rightarrow\left\{\begin{array}{lr}p(1-\alpha)\,,\;&\;\,p\leq
\alpha\\\alpha(1-p)\,,\;&\; p\geq \alpha\end{array}\right.\label{fl}\,.
\end{equation}
In this limit,
 \begin{eqnarray}
S_f(\rho)&=&{\rm Tr}[g_t(\rho-\alpha)]-(D-1)g_t(-\alpha)-g_t(1-\alpha)
\nonumber\\&\rightarrow& \alpha(n_\alpha-1)+1-\sum_{p_i>\alpha}p_i\,,
 \end{eqnarray}
where $D=d_A d_B$ is the dimension of $\rho$ and $n_\alpha$ is the number of 
eigenvalues of $\rho$ larger than $\alpha$. It is now seen that if 
${\rho\prec\!\!\!\!\!/\,\rho_A}$, $S_f(\rho)<S_f(\rho_A)$ for some $\alpha$ 
and $t$. Suppose that the inequalities (\ref{m}) are first violated for $i=j$, 
i.e., ${\cal S}_j>{\cal S}^A_j$, which implies $p_j>p^A_j$. Choosing 
$\alpha=p^{A}_j$, we have $n_\alpha(\rho)=j'\geq j$, obtaining, for 
$t\rightarrow \infty$,
 \begin{eqnarray}
S_f(\rho)-S_f(\rho_A)&\rightarrow &p^A_j(j'-j)-({\cal S}_{j'}-{\cal S}^{A}_{j})
\nonumber\\
&\leq& -[(p_{j'}-p^{A}_j)(j'-j)+{\cal S}_j-{\cal S}^{A}_j]<
0\,.\nonumber
 \end{eqnarray}
Hence $S_f(\rho)<S_f(\rho_A)$ for sufficiently large $t$ and $\alpha$
sufficiently close to $p^A_j$.

In the opposite limit $t\rightarrow 0^+$, and assuming $g_t(x)=h(tx)/t$, as in
Eq.\ (\ref{gtx}), $f(\rho)\rightarrow t|h''(0)|(\rho-\rho^2)/2$, becoming hence
independent of $\alpha$ and proportional to the Tsallis case $q=2$. On the
other hand, for $\alpha=0$ or $1$ and $t\rightarrow\infty$, the lowest non-zero
order of $f$ depends on the higher order term $r_t(x)=g_t(x)+|x|/2-g_t(0)$. For
instance, if $\alpha=1$, $f(\rho)=r_t(\rho-1)-(1-\rho)r_t(-1)$, with
$r_t(x)=-(2t)^{-1}\ln[(1+e^{-2t|x|})/2]$ in the case (\ref{gtx}).

\begin{figure}[t]
\vspace*{-3cm}

\hspace*{-2cm}\scalebox{0.7}{\includegraphics{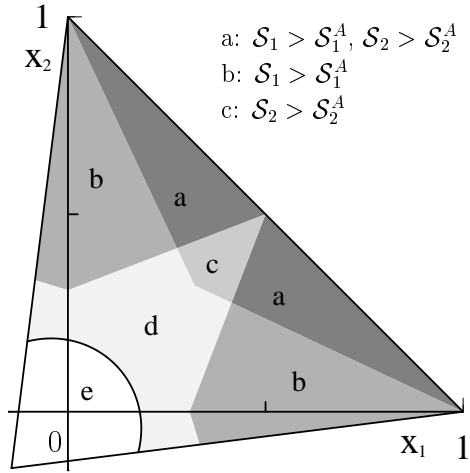}}\vspace*{-10.75cm}

\caption{Region of allowed values of $x_1$, $x_2$ in Eq.\ (\ref{r1}), for
$d=3$. $\rho$ is entangled in sectors $a,b,c,d$, and separable in $e$. The
disorder criterion detects entanglement just in sectors $a,b,c$, where the
inequalities (\ref{m}) that are not fulfilled are indicated.}
\label{f2}\end{figure}

{\it Example}. Let us consider a system of two qudits ($d$-dimensional
subsystems) with $d\geq 3$. We will first examine a mixture of two
antisymmetric states with a full random state,
\begin{equation}
\rho=x_1|01^-\rangle\langle 01^-|+x_2|02^-\rangle\langle 02^-|+ yI_A\otimes I_B
\,,\label{r1}
\end{equation}
where $|ij^-\rangle=(|ij\rangle-|ji\rangle)/\sqrt{2}$, $y=(1-x_1-x_2)/d^2$ and
$I_{A,B}=\sum_{i=0}^{d-1}|i\rangle\langle i|$. For $x_2=0$ and $d=2$, it
reduces to a Werner-Popescu state for two qubits \cite{W.89,Po.94}. Its
eigenvalues are obviously $x_1+y$, $x_2+y$ and $y$ ($(d^2-2)$-fold degenerate).
Positivity of $\rho$ imposes then the conditions $x_1+x_2\leq 1$, $x_i\leq
(d^2-1)x_j+1$, $i\neq j$, which determine a triangle $R$ with vertices at
$\bm{x}\equiv (x_1,x_2)=(1,0)$, $(0,1)$, $-(1,1)/(d^2-2)$, shown in Fig.\
\ref{f2} for $d=3$. Negative values of $x_i$ are in principle feasible and
represent a depletion of the antisymmetric states with respect to the fully
mixed state.

The lowest eigenvalue of the partial transpose of $\rho$ is
\begin{equation}
\sigma(\bm{x})=y-|\bm{x}|/2\,,\label{pc}
\end{equation}
where $|\bm{x}|=(x_1^2+x_2^2)^{1/2}$. The equation $\sigma(\bm{x})=0$
determines an ellipse centered at $x_1=x_2=-4/(d^4-8)$, with axes rotated 45
degrees with respect to the $x_1,x_2$ axes, and $\sigma(\bm{x})<0$ outside this
ellipse. Hence, according to the Peres criterion \cite{P.96}, $\rho$ will be 
entangled in the intersection of this region with $R$ (see Fig.\ \ref{f2}).
Moreover, it can be easily shown that $\rho$ is indeed separable if
$\sigma(\bm{x})\geq 0$. We may rewrite (\ref{r1}) in the form
 \begin{eqnarray}
\rho&=&\sum_{i=1}^n\rho_i+y(I_A\otimes
I_B-\sum_{i=1}^nQ_i)\,,\;\;\;\;\label{rhoi}
\\\rho_i&\equiv &x_i|0i^-\rangle\langle 0i^-|+y Q_i\,,\\
Q_i&=&q_i|00\rangle\langle 00|+|0i\rangle\langle 0i|+|i0\rangle\langle i0|
+|ii\rangle\langle ii|\label{Pi}\,,
 \end{eqnarray}
where $n=2$, $q_i\geq 0$, $q_1+q_2=1$. For $x_1,x_2\in R$, $\rho_i$ is a
non-negative operator for a two qubit system, where the Peres criterion is 
sufficient \cite{HHH.96}, and will then be separable if and only if the lowest 
eigenvalue of its partial transpose is non-negative. This leads to the 
condition 
 \begin{equation}
 x_i^2\leq 4q_iy^2\,.\label{xi}
 \end{equation}
If Eq.\ (\ref{xi}) holds for $i=1,2$, then  $|\bm{x}|^2\leq 4y^2$, implying
$\sigma(\bm{x})\geq 0$. Moreover, choosing $q_i=x_i^2/|\bm{x}|^2$, it is seen
that {\it any} state satisfying $\sigma(\bm{x})\geq 0$ will fulfill Eqs.\
(\ref{xi}) for $i=1,2$, implying separability of $\rho_i$ and hence of $\rho$.
Thus, $\rho$ is entangled if and only if $\sigma(\bm{x})<0$. Denoting with
$\gamma$ the angle between $\bm{x}$ and ${\bm v}=(1,1)$, this implies
entanglement for 
\begin{equation} 
 |\bm{x}|>[\sqrt{2}\cos\gamma+d^2/2]^{-1}\,.\label{rc}
 \end{equation}
The entanglement threshold for $|\bm{x}|$ is minimum at $\gamma=0$ ($x_1=x_2$),
and decreases as $d^{-2}$ as $d$ increases (for $d=3$ the entangled region
already covers $87\%$ of $R$). If $x_2=0$ Eq.\ (\ref{rc}) implies entanglement
just for
\begin{equation}
x_1>(1+d^2/2)^{-1}\,.\label{rc1}
 \end{equation}
For $d=2$ this yields  $|x_1|>1/3$, in agreement with the well-known result for
a Werner-Popescu state \cite{P.96,B.96}.

As seen in Fig.\ \ref{f2}, the disorder criterion detects entanglement only in
a sector of the entangled region (covering $77\%$ of it for $d=3$), which can
be classified according to the subset of inequalities (\ref{m}) that are
violated. Note that the complement of this sector is not convex. The reduced
density matrix of any of the subsystems is
\[
\rho_A=\frac{x_1+x_2}{2}|0\rangle\langle 0|+\frac{x_1}{2}|1\rangle\langle 1|+
\frac{x_2}{2}|2\rangle\langle 2|+y\, d\, I_A\,.
\]
Hence, for $x_1\geq x_2\geq 0$, ${\cal S}_1>{\cal S}_1^A$ if
$x_1+y>(x_1+x_2)/2+y d$, i.e.,
\begin{equation}
x_1>\frac{1+x_2(\delta-1)}{\delta+1}\,,
\;\;\;\delta\equiv\frac{d^2}{2(d-1)}\,,\label{a}
\end{equation}
while if $x_1\geq 0\geq x_2$, this occurs for $x_1+y>x_1/2+y d$, i.e.,
$x_1>(1-x_2)/(1+\delta)$. Besides, for $x_1\geq x_2$, ${\cal S}_2>{\cal S}_2^A$
in $R$ just for $x_1+x_2+2y>x_1+x_2/2+2yd$, i.e.,
\begin{equation}
x_1>1-x_2(1+\delta/2)\,.\label{c}
\end{equation}
These regions partially overlap, giving rise, together with their symmetric
partners for $x_2>x_1$, to the  sectors $a,b,c$ of Fig.\ \ref{f2}. In $c$, only
the second inequality is violated. For $3\leq i\leq d$, ${\cal S}_i\leq {\cal
S}^{A}_i$ $\forall$ $x_1,x_2\in R$, with ${\cal S}_i={\cal S}_i^A$ only along
the border $x_1+x_2=1$ (where ${\cal S}_i=1$ for $i\geq 2$).

In particular, for $x_2=0$ the disorder criterion predicts
entanglement just for
\begin{equation}
x_1>(1+\delta)^{-1}\,,\label{rc3}
\end{equation}
which provides a bound that is strictly larger than Eq.\ (\ref{rc1}) except for
$d=2$, and decreases only as $d^{-1}$ as $d$ increases. The same occurs for
$x_1=x_2$ ($\gamma=0$), where Eq.\ (\ref{c}) predicts entanglement just for
$|\bm{x}|>[\sqrt{2}+\delta/(2\sqrt{2})]^{-1}$.

\begin{figure}[t]
\vspace*{-3cm}

\hspace*{-2cm}\scalebox{0.7}{\includegraphics{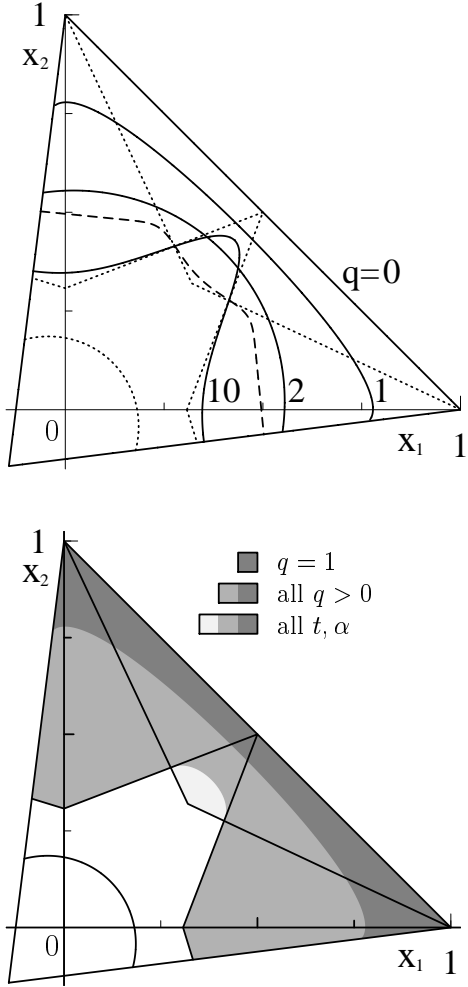}} \vspace*{-3.75cm}

\caption{Top: Curves where $S_f^A(\rho)=0$ according to: the Tsallis entropy
for the indicated values of $q$ (solid lines), and the entropy (\ref{gt})
(dashed line) with $g$ given by Eq.\ (\ref{gtx}) and $\alpha=0.28$, $t=30$. The
curve for $q=1$ corresponds to the von Neumann entropy. Dotted lines indicate
the boundaries of the sectors of Fig.\ \ref{f2}. Bottom: Union of sectors where
$S_f^A(\rho)<0$ according to: the von Neumann entropy ($q=1$), the whole set of
Tsallis entropies (all $q>0$) and the whole set of entropies
(\ref{gt}-\ref{gtx}) ($t>0$, $0<\alpha<1$).}
 \label{f3}\end{figure}

For any entropic function $f$, the region where Eq.\ (\ref{Sf1}) does not hold,
i.e., where
\begin{equation}
S_f^A(\rho)\equiv S_f(\rho)-S_f(\rho_A)\,,\label{SA}
\end{equation}
is negative, will be necessarily contained in sectors $a,b,c$. As seen in Fig.\ 
\ref{f3} for $d=3$, this region is rather small for the von Neumann entropy
($q=1$), but in the Tsallis case it increases as $q$ increases, covering
sectors $a$ and $b$ for $q\rightarrow\infty$, although it never covers sector
$c$. Even the region where $S_f^A(\rho)<0$ for {\it some} $q>0$, depicted in
the bottom panel, leaves out an appreciable fraction of sector $c$. The border
of this region in sector $c$  corresponds to {\it finite} and varying values of 
$q$, which can be determined by the simultaneous conditions $S_f^A(\rho)=0$,
$\partial S_f^A(\rho)/\partial q=0$. Note also that along the outer border
($x_1+x_2=1$), $S_f^A(\rho)\leq 0$ for {\it any} $f$ since here 
$\rho^A\prec\rho$. As $q\rightarrow 0^+$, the region where $S_f^A(\rho)<0$ 
shrinks and approaches this line. 

\begin{figure}[t]
\vspace*{-2cm}

\hspace*{-2cm}\scalebox{0.65}{\includegraphics{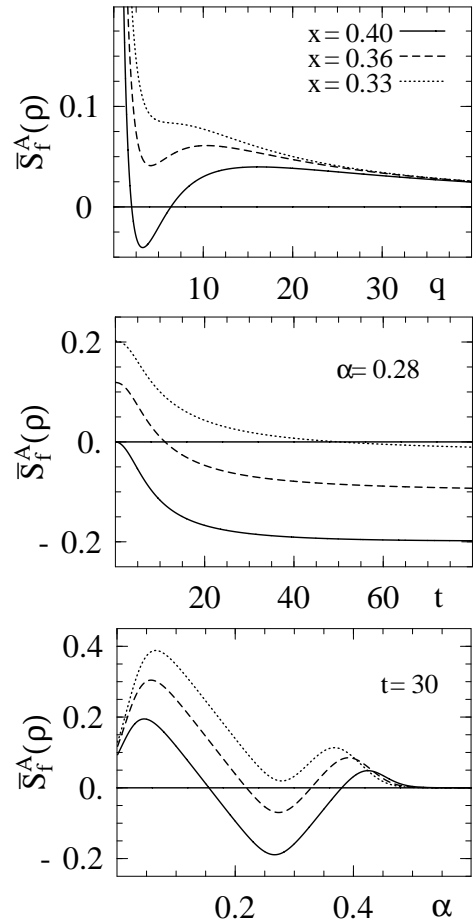}} \vspace*{-5.cm}

\caption{The behavior of $\bar{S}_f^A(\rho)$ for $x_1=x_2=x$ (sector c of Fig.\
\ref{f2}) at the indicated values of $x$, according to: the Tsallis entropy as
a function of $q$ (top), and the entropy (\ref{gt}) for the case (\ref{gtx}),
as a function of $t$ (center) and $\alpha$ (bottom). }
\label{f4}\end{figure}

The typical behavior of the Tsallis conditional entropy
$\bar{S}_f^A(\rho)\equiv S_f^A(\rho)/{\rm Tr}[\rho_A^q]$ in sector $c$ is
depicted in the top panel of Fig.\ \ref{f4} as a function of $q$. Here
$S_f^A(\rho)$ may become negative only in a finite interval of $q$ values. For
$x_1=x_2=x$, this occurs just for $x>0.381$ (at this point $S_f^A(\rho)=0$ just
for $q=3.575$). In comparison, along this line the disorder criterion predicts
entanglement for $x>(2+\delta/2)^{-1}=0.32$, while the true value given by the 
Peres criterion is $x>(2+d^2/\sqrt{2})^{-1}\approx 0.12$ (in the von Neumann
case, $S_f^A(\rho)<0$ just for $x> 0.452$). Note as well that
$\bar{S}_f^A(\rho)$ does not necessarily develop a local minimum as a function
of $q$ in this sector (for $x_1=x_2=x$ this occurs just for $x>0.334$).

In the case of the functions (\ref{gt}), $S_f^A(\rho)$ becomes negative for
sufficiently large $t$ at any point of sectors $a,b,c$ if $\alpha$ is
appropriately chosen. Due to the symmetries of the present case, just two
values of $\alpha$ are required: For $d=3$, in sectors $a,b$ it is sufficient
to choose {\it any} single $\alpha>5/13\approx  0.385$, which is the value of
$p_1^A$ at the tip of sector $b$ ($\bm{x}=(4/13,0)$), whereas the whole sector
$c$ can be covered choosing $\alpha=7/25=0.28$, which is the value of $p_2^A$
at the inner tip of this sector ($x_1=x_2=0.32$). In the case (\ref{gtx}),  the
curves where $S_f^A(\rho)=0$ are, for large $t$ and $\alpha>1/2$, similar to
those of the Tsallis case for large $q$, whereas if $\alpha=0.28$, they
approach for large $t$ the inner border of sector $c$, as seen  in the top
panel of Fig.\ \ref{f3}. The behavior of $\bar{S}_f^A(\rho)\equiv
S_f^A(\rho)/|{\rm Tr}[g_t(\rho)]|$ in sector $c$ is depicted in Fig.\ \ref{f4},
where it is seen that for sufficiently large $t$, it becomes negative in a {\it 
finite} interval of $\alpha$ values (which always includes $\alpha=0.28$), for
any $x>0.32$.

Similar considerations hold for the states
\begin{equation}
\rho=\sum_{i=1}^nx_i|0i^-\rangle\langle 0i^-|+y I_A\otimes I_B\,, \label{rh2}
\end{equation}
where $n\leq d-1$  and $y=(1-\sum_{i=1}^nx_i)/d^2$. Its eigenvalues are
$x_i+y$, $i=1,\ldots,n$, and $y$ ($(d^2-n)$-fold degenerate), and the allowed
values of $x_i$ are contained in a region $R$ bounded by the conditions $y\geq
0$, $x_i\geq -y$.

The lowest eigenvalue of the partial transpose is again given by Eq.\
(\ref{pc}),  with $|\bm{x}|=[\sum_{i=1}^nx_i^2]^{1/2}$, and it can be seen that
$\rho$ is entangled if and only if $\sigma(\bm{x})<0$: writing $\rho$ in the
form (\ref{rhoi}), with $q_i\geq 0$, $\sum_{i=1}^nq_i=1$, {\it any} state
satisfying $\sigma(\bm{x})\geq 0$ is seen to fulfill Eqs.\ (\ref{xi}) for
$i=1,\ldots,n$ if $q_i=x_i^2/|\bm{x}|^2$, which implies separability of
$\rho_i$ and hence, of $\rho$. The entanglement condition $\sigma(\bm{x})<0$
can be cast as
\begin{equation}
|\bm{x}|>(\sqrt{n}\cos\gamma+d^2/2)^{-1}\label{rcg}\,,
 \end{equation}
where $\gamma$ is the angle between $\bm{x}$ and $\bm{v}=(1,\ldots,1)$, and
corresponds to the region outside an $n$-dimensional ellipsoid. For
$|\gamma|<\pi/2$ and fixed $d$, the entanglement threshold for $|\bm{x}|$
decreases as $n$ increases.

\begin{figure}[t]
\vspace*{-3cm}

\hspace*{-2.5cm}\scalebox{0.75}{\includegraphics{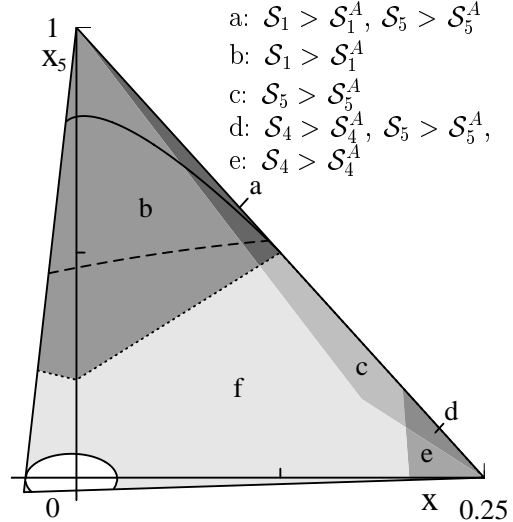}} \vspace*{-11.5cm}

\caption{Section of allowed values of $x,x_5$ in the state (\ref{rh2}) for
$d=6$, $n=5$ and $x_i=x$ for $i=1,\ldots,4$. $\rho$ is entangled in all shaded
sectors. The disorder criterion detects entanglement in sectors $a,b,c,d,e$,
where the inequalities (\ref{m}) that are not fulfilled are indicated. Also
depicted are the curves where $S_f^A(\rho)=0$ in the von Neumann case (solid
line) and the Tsallis case $q=2$ (dashed line), such that $S_f^A(\rho)<0$ above
these curves. The dotted line indicates the boundary of the region covered by
all $q>0$ in the Tsallis case (see text).}
\label{f5}\end{figure}

For $n\geq 5$ (and $d\geq n+1$) a novel feature arises: entanglement may also
occur for $x_i<0$ $\forall$ $i$, i.e., in the vicinity of $\gamma=\pi$, since
the vertex of $R$ in this direction ($x_{vi}=-(d^2-n)^{-1}$ for $i=1,\ldots,n$)
lies outside the ellipsoid: $|\bm{x}_v|=\sqrt{n}/(d^2-n)>(d^2/2-\sqrt{n})^{-1}$
for $n>4$. Thus, for $n\geq 5$ entanglement can also occur by {\it depletion}
of antisymmetric states, although the corresponding region is very small (see
Fig.\ \ref{f5} for the case $n=5$, $d=6$). Nevertheless, the ellipsoid
$\sigma(\bm{x})\geq 0$ is never fully contained in $R$. Points of the boundary
of $R$ like $x_j=0$ if $j\neq i$, $x_i=-1/(d^2-1)$, fulfill $\sigma(\bm{x})>0$
$\forall$ $d\geq 2$.

The disorder criterion ensures again entanglement in a much smaller region.
Moreover, the first violation of the inequalities (\ref{m}) may now occur for
{\it any} $i$ between $1$ and $n$. The reduced density of system $A$ is
 \begin{equation}
\rho_A=\frac{1}{2}\sum_{i=1}^nx_i[|i\rangle\langle i|+|0\rangle\langle 0|]
 +y d I_A\,.
 \end{equation}
For $x_1\geq x_2\geq\ldots \geq x_n\geq 0$ and $1\leq i\leq n$,
 ${\cal S}_i>{\cal S}_i^A$ for
\[x_i>\frac{1}{2}\sum_{j=i}^{n}x_j+iy(d-1)\,.\]
According to the values of the $x_i$, we may have violation of the inequalities
(\ref{m}) for {\it any} set of $m$ distinct indices $i_j$, with $m\leq n$,
$i_j\leq n$. In particular, if $x_i=x\geq 0$ for $i=1,\ldots,n$ ($\gamma=0$),
only the $n^{\rm th}$ inequality can be violated: ${\cal S}_n>{\cal S}^A_n$ for
 \begin{equation}
x>n/(n^2+\delta)\,,\label{betn}
 \end{equation}
i.e., $|\bm{x}|>[\sqrt{n}+\delta/n^{3/2}]^{-1}$. This bound is larger than that
given by Eq.\ (\ref{rcg}) for $\gamma=0$, for any $n\geq 1$, $d\geq 3$. On the
other hand, the disorder criterion is not able to predict entanglement for
$x<0$ ($\gamma=\pi$) for any $d$, $n$.

For instance, Fig.\ \ref{f5} depicts, for $n=5$ and $d=6$, the intersection of
$R$ with the plane $x_i=x$ for $i=1,\ldots,4$. The separable sector covers only
a small fraction of this section ($\sim 2,6\%$) but the disorder criterion
covers only $\approx 51\%$ of the entangled region. Moreover, the first
violation of the inequalities (\ref{m}) may  occur for $i=1$, 4 or 5, the last
two in sectors $c,d,e$, although the largest region corresponds to $i=1$
(sectors $a,b$, accounting for $40\%$ of the entangled region).

The region where $S_f^A(\rho)<0$ is small in the von Neumann case, but
increases as $q$ increases in the Tsallis case,  covering sectors $a$ and $b$
for $q\rightarrow \infty$, although the covering of sectors $c,d,e$ is very
poor. The inner border of the region where $S_f^A(\rho)<0$ for some $q>0$ lies
very close to the outer boundary ($\sum_{i=1}^nx_i=1$) in these sectors, being
indistinguishable from it in the scale of the figure. For instance, for $x_5=x$
(sector $c$), the boundary lies at $x=0.2$ and $S_f^A(\rho)<0$ for some $q$
just for $x>0.19997$, while for $x_5=0$ (sector $e$), the boundary is at
$x=0.25$ and $S_f^A(\rho)<0$ only for $x>0.2492$. In comparison, along these
lines the disorder criterion predicts entanglement for $x>0.175$ and $x>0.204$
respectively, as given by Eq.\ (\ref{betn}). For the entropy (\ref{gt}), it is
necessary to employ in this case an {\it interval} of $\alpha$ values in order
to have $S_f^A(\rho)<0$ for large $t$ in all points of regions $c,d,e$
($\alpha\in[0.069,0.108]$ in $c$ and $\alpha\in[0.114,0.134]$ in $e$), although 
for the whole sectors $a$ and $b$ it is sufficient to take a single
$\alpha>1/4$. It should be mentioned that all previous equations and results
remain valid if the antisymmetric states in (\ref{r1}) or (\ref{rh2}) are
replaced by symmetric states $|ij^+\rangle=(|ij\rangle+|ji\rangle)/\sqrt{2}$.

In conclusion, we have provided an example of a set of smooth entropic forms
which are always able to detect the breakdown of the majorization relations
(\ref{m}) in entangled states, through the sign of the entropic difference
(\ref{SA}). We have also shown that when the first violation occurs for $i>1$,
the Tsallis conditional entropy may not detect such violations for any $q$,
being hence weaker in these cases than the disorder criterion. The present
entropic forms are therefore necessary for detecting and measuring entanglement
by entropic means in such situations. These issues have been illustrated in
detail for particular states of a two qudit system, which show that the first
violation of Eqs.\ (\ref{m}) may in fact occur for any $i\leq d-1$.

RR and NC acknowledge support from CIC and CONICET, respectively, of Argentina,
and a grant of Fundaci\'on Antorchas.

\end{document}